\documentclass[a4paper,11pt]{article}
\usepackage{pos}

\title{Lattice Field Theory for a network of real neurons}

\author*[a,b]{Simone Franchini}
\author[b]{Giampiero Bardella}

\affiliation[a]{CNR--ISTC, Via Gian Domenico Romagnosi 18, 00196 Rome, Italy}

\affiliation[b]{Department of Physiology and Pharmacology, Sapienza University of Rome, Piazza Aldo Moro 1, 00185 Rome, Italy}

\emailAdd {simone.franchini@yahoo.it}
\emailAdd {giampiero.bardella@uniroma1.it}

\abstract{In a recent paper [Bardella et al., Entropy \textbf{26} (6), 495 (2024)] we introduced a simplified Lattice Field Theory (LFT) framework that allows experimental recordings from major Brain--Computer Interfaces (BCIs) to be interpreted in a simple and physically grounded way.
From a neuroscience point of view, our method modifies the Maximum Entropy model for neural networks so that also the time evolution of the system is taken into account and it can be interpreted as another version of the Free Energy principle (FEP).
This framework is naturally tailored to interpret recordings from chronic multi--site BCIs, especially spike rasters from measurements of single neuron activity.}

\begin{document}

\makeatletter
\def\logo{}             
\def\@journal{}         
\def\@PoScopyright{}    
\def\PoScopyright{}     
\def\PoS@copyright{}    
\makeatother

\maketitle
\thispagestyle{empty}
\newpage

\section{Introduction}

\noindent{}
In \cite{Bardella,iScience2024,Franchini2023} we introduce a simplified
\textbf{Lattice Field Theory} (LFT) framework that allows experimental
observations from the most popular Brain--Computer Interfaces (BCIs)
to be interpreted in a simple and physically grounded way, and that
can also deal with the dynamics of the system.
Although several proposals for a \textbf{Neural LFT} already exist (for
example \cite{Buice,Chow,Gosselin,Halverson,Fagerholm}),
in our opinion none tackled the problem of connecting with the experimental observables
in such a way that it could be managed also by "non--physicists".
Therefore,
in \cite{Bardella,iScience2024,Franchini2023} we elaborated a minimalistic
formulation of a Neural LFT that does not rely on the common particle
physics background, which considerably simplify the arguments, and
also provides a framework to handle both biological and artificial
neural networks on a common ground.
The framework was developed primarily
to evaluate the population dynamics of single unit (neuron) brain
activity recorded from arrays of multiple electrodes \cite{Pani},
and interprets the relation between microscopic parameters and
experimental observations in terms of a renormalization by decimation
\cite{Bardella,iScience2024}.
\section{Why a LFT?}

\noindent{}
The \textbf{Max Entropy} model for biological neurons was introduced by Schneidman, Tkacik, Bialek et al.
in \cite{Schneidman,Tkacick,Meshulam} about twenty years ago and it consists in fitting the neuron--neuron correlation matrix (averaged on a certain time interval) with the Ising model.
From the neuroscience perspective this also can be interpreted as a kind of Free Energy principle (FEP, see in Section 4).
A fundamental limitation of \cite{Schneidman,Tkacick,Meshulam} is that it can deal only with stationary activities.
As explained in Section V.B of the recent review \cite{Meshulam} by Meshulam and Bialek (page 16): "Going beyond stationary situations to building these extended models once seemed impossible, but the push for stable recordings of electrical activity overs days and even weeks [...] will create new opportunities".
Our NLFT \cite{Bardella} does exactly this: it extends the Max Entropy model to include time evolution by embedding not in the framework of Statistical Mechanics (SM) \cite{Schneidman,Tkacick,Meshulam} but in that of Quantum Mechanics.
The crucial advantage is best explained in the context of, e.g., the Parisi--Wu quantization method \cite{ParisiWu,StocQuant}: the key is that the time evolution is naturally included in the theory.
In fact, although the analogy between Euclidean QFT and the canonical ensemble establishes a common ground for the mathematical and computational techniques in the SM framework like \cite{Schneidman,Tkacick,Meshulam} the time evolution is used to define the statistical averages.
In a quantum system of Parisi--Wu type the quantization is achieved along a different "fictional" time direction so that the original time remains at our disposal to deal with the non--stationary situations.
Therefore the QFT framework of \cite{Bardella} (or equivalent) is mandatory to deal with non--stationary time evolution and the SM framework of \cite{Schneidman,Tkacick,Meshulam} can be seen as its stationary limit (see Section 2.2 of \cite{Bardella}).
\section{Neural LFT}

\noindent{}
We postulate that the evolution of a network of neurons can be represented
by a discrete process of interacting binary fields, or "\textbf{qubits}"
\cite{Weizsacker,LLoyd,PerettoQCD,Deutsch,Franchini2023}.
More precisely, we will assume that from the functional point of view the
neuron can be represented by a binary variable and that a system of
neurons can be described by a string of bits \cite{Schneidman,Peretto}, like in a Turing machine.
Let us introduce the following notation for the (binary) support of the neuron state:
\begin{equation}
\Gamma:=\left\{ 0,1\right\} 
\end{equation}
and let $N$ be the number of neurons involved in a given task, these
are mapped on the set of vertices
\begin{equation}
V:=\left\{ 1\leq i\leq N\right\} .
\end{equation}
Let $T$ be the registered time--span in units of the \textbf{absolute refractory
period} \cite{Bardella}.
We can slice that interval into blocks of that size and map
them onto the vertex set
\begin{equation}
S:=\left\{ 1\leq\alpha\leq T\right\} .
\end{equation}
The previous considerations already imply that when studying a system of
neurons we can map our space--time analogue on a discrete set of lattice sites.
Let $\Omega_i^\alpha$ be the state of the $i$--th neuron at time $\alpha$, we establish the following notation (\textbf{kernel
representation}, \cite{Franchini2023, Franchini2021,FranchiniSPA2023,Franchini2024,FranchiniCavity2026,FranchiniMulti2025}) for the raster: 
\begin{equation}
\Omega:=\left\{ \,\Omega_{i}^{\alpha}\in\Gamma:\alpha\in S,\ i\in V\right\} \in\Gamma^{NT}.
\end{equation}
The size (cardinality) of the allowed configuration space is, therefore,
$2^{NT}$.
The neural dynamics is expected to follow a time evolution, where the
state at given time is causally influenced by previous states.
As already argued by many
authors \cite{Buice,Chow,Gosselin,Halverson,Fagerholm},
it is reasonable to assume that such dynamics can be described (at least
formally) by some quantum evolution, so that the apparatus of \textbf{quantum
field theory} \cite{Guerra,ParisiWu,StocQuant,Parisi},
and especially that of LFT \cite{Wilson,Lee,PerettoQCD,Weise},
can be applied.
Mimicking the Landau approach to classical mechanics \cite{Landau} we postulate
ab initio the existence of the \textbf{Euclidean action} function \cite{Guerra,ParisiWu,StocQuant}, denoted by the symbol
\begin{equation}
\mathscr{A}:\Gamma^{NT}\rightarrow\mathbb{R}\,.
\end{equation}
We introduce the analogue partition function and the Gibbs measure:
\begin{equation}
Z:=\sum_{\Omega\in\Gamma^{NT}}\exp\left[-\hslash^{-1}\mathscr{A}\left(\Omega\right)\right],\ \ \ \mu\left(\Omega\right):=\frac{\exp\left[-\hslash^{-1}\mathscr{A}\left(\Omega\right)\right]}{Z}.
\end{equation}
The ensemble average of the generic observable is
\begin{equation}
\langle\mathscr{O}\rangle_{\mu}:=\sum_{\Omega\in\Gamma^{NT}}\mathscr{O}\left(\Omega\right)\mu\left(\Omega\right).
\end{equation}
The combined efforts of many authors (see, e.g., \cite{Guerra,ParisiWu,StocQuant})
showed that this ensemble average is actually equal to the quantum
average.
The non--quantum regime is attained when $\hslash\rightarrow0$,
and it is therefore identified with the \textbf{ground state} of the action.
Notice that in this limit case the dynamics becomes conservative (in the sense that there are no dissipative dynamics in place) and so the path pursued by the system is always that of least action.
\section{Relation with the Free Energy principle}

\noindent{}
The link with neuroscience is provided by the Free Energy principle (FEP,   \cite{Friston1,Friston2,Friston3}).
Let us multiply and divide the weights of the partition function by a test measure $\zeta$
\begin{multline}
Z:=\sum_{\Omega\in\Gamma^{NT}}\exp\left[-\hslash^{-1}\mathscr{A}\left(\Omega\right)\right]=\\
=\sum_{\Omega\in\Gamma^{NT}}\zeta\left(\Omega\right)\,\exp\left[-\hslash^{-1}\mathscr{A}\left(\Omega\right)-\log\zeta\left(\Omega\right)\right]=\\
=\langle\exp\left[-\hslash^{-1}\mathscr{A}\left(\Omega\right)-\log\zeta\left(\Omega\right)\right]\rangle_{\zeta}\label{eq:zorn}
\end{multline}
applying \textbf{Jensen inequality} to the exponential we immediately find
\begin{equation}
\langle\exp\left[-\hslash^{-1}\mathscr{A}\left(\Omega\right)-\log\zeta\left(\Omega\right)\right]\rangle_{\zeta}\geq\exp\left[-\hslash^{-1}\langle\mathscr{A}\left(\Omega\right)\rangle_{\zeta}-\langle\,\log\zeta\left(\Omega\right)\rangle_{\zeta}\right].
\end{equation}
We can easily recognize the canonical Free Energy functional in the exponential on the right side
\begin{equation}
\mathcal{F}\left(\zeta\right):=\langle\mathscr{A}\left(\Omega\right)\rangle_{\zeta}+\hslash\,\langle\,\log\zeta\left(\Omega\right)\rangle_{\zeta}.
\end{equation}
Therefore, for any trial distribution $\zeta$ holds the inequality
\begin{equation}
-\hslash\log Z\leq\mathcal{F}\left(\zeta\right),\ \ \ \forall\zeta\in\mathcal{P}\,(\Gamma^{NT}).
\end{equation}
The minimum is attained by the Gibbs measure $\mu$
\begin{equation}
\mathcal{F}\left(\mu\right)=\inf_{\zeta\in\mathcal{P}\,(\Gamma^{NT})}\mathcal{F}\left(\zeta\right)
\end{equation}
and is exactly the Free Energy (Free Action in our context) 
\begin{equation}
\mathcal{F}\left(\mu\right)=-\hslash \log Z.
\end{equation}
The FEP provides the connection with Bayesian Brain theories.
For example is the pivotal concept of \textbf{Active Inference}, a process theory that aims to explain the interactions between any agent and its environment by first principles \cite{Friston1,Friston2,Friston3}.
The scope of the agent is to minimize the discrepancy between predictions and observations, either by adapting the environment to the predictions or the predictions to the environment \cite{Friston1,Friston2,Friston3}.
The measure of such discrepancy is identified in the Free Energy functional by interpreting the partition function with some (un--normalized) Bayesian model evidence (see Chapter 4.2 of \cite{Friston3}).
\section{Taylor expansion of the action and two--body truncation}

\noindent{}
By Taylor\textquoteright s theorem, the action can be expanded as
follows:
\begin{multline}
\mathscr{A}(\Omega|\,F,I):=\sum_{i\in V}\sum_{\alpha\in S}I_{i}^{\alpha}\Omega_{i}^{\alpha}+\sum_{i\in V}\sum_{j\in V}\sum_{\alpha\in S}\sum_{\beta\in S}F_{ij}^{\alpha\beta}\Omega_{i}^{\alpha}\Omega_{j}^{\beta}+\\
+\sum_{i\in V}\sum_{j\in V}\sum_{h\in V}\sum_{\alpha\in S}\sum_{\beta\in S}\sum_{\gamma\in S}F_{ijh}^{\alpha\beta\gamma}\Omega_{i}^{\alpha}\Omega_{j}^{\beta}\Omega_{h}^{\gamma}+\\
+\sum_{i\in V}\sum_{j\in V}\sum_{h\in V}\sum_{k\in V}\sum_{\alpha\in S}\sum_{\beta\in S}\sum_{\gamma\in S}\sum_{\delta\in S}F_{ijhk}^{\alpha\beta\gamma\delta}\Omega_{i}^{\alpha}\Omega_{j}^{\beta}\Omega_{h}^{\gamma}\Omega_{k}^{\delta}+\,...\label{eq:cvcg}
\end{multline}
The terms are the one--, two--, three--, and four--vertex interactions
etc., while the tensors $F$ collects the \textbf{parameters of the
theory}.
The number of parameters
to describe the general $n-$vertices theory \cite{FranchiniMulti2025} would be $N^{n}T^{n}$, but here we postulate (part because the correlations observed in \cite{Bardella}
and \cite{Schneidman} are small and part due to possible computational limits) that the terms with more than two vertices can be neglected,
which result in the following simplification of the parameter tensor:
\begin{equation}
F_{ijh}^{\alpha\beta\gamma}=0,\ \ \ F_{ijhk}^{\alpha\beta\gamma\delta}=0,\ \ \ ...
\end{equation}
Therefore, the proposed action reduces to:
\begin{equation}
\mathscr{A}(\Omega|\,F,I)=\sum_{i\in V}\sum_{\alpha\in S}I_{i}^{\alpha}\Omega_{i}^{\alpha}+\sum_{i\in V}\sum_{j\in V}\sum_{\alpha\in S}\sum_{\beta\in S}F_{ij}^{\alpha\beta}\Omega_{i}^{\alpha}\Omega_{j}^{\beta}.
\end{equation}
Notice that this is formally equivalent to the Max Entropy model applied by
Schneidman et al.
2006 \cite{Schneidman}, the crucial difference
is that here \textbf{the same neuron at different times is considered
like two different neurons}, is the action that ultimately makes them
look the same evolving in time.
Let us introduce the observable "grand covariance" \cite{Bardella}:
\begin{equation}
\mathcal{C}_{ij}^{\,\alpha\beta}:=\langle\,\Omega_{i}^{\alpha}\Omega_{j}^{\beta}\rangle_{\mu}-\langle\,\Omega_{i}^{\alpha}\rangle_{\mu}\langle\,\Omega_{j}^{\beta}\rangle_{\mu}
\end{equation}
from which the parameters can be inferred 
\begin{equation}
\{F,\,I\}\rightleftharpoons\{\mathcal{C},\langle\Omega\rangle_{\mu}\}.
\end{equation}
The ability of reconstructing the couplings has become a major
goal of computational neuroscience in the last decades, and powerful
inference methods are now available, see \cite{Berg} for a review.
\section{Causality, local memory and bi--stationarity of the couplings}

\noindent{}
We can further simplify by implementing the causal constraint
respect to the time variable $\alpha$
\begin{equation}
F_{ij}^{\alpha\beta}=0,\ \ \ \beta>\alpha\,.
\end{equation}
We now separate the links with $\beta=\alpha$ associated to the space--like
interactions,
\begin{equation}
\mathscr{A}(\Omega|\,F,I)=\sum_{i\in V}\sum_{\alpha\in S}I_{i}^{\alpha}\Omega_{i}^{\alpha}
+\sum_{i\in V}\sum_{j\in V}\sum_{\alpha\in S}F_{ij}^{\alpha\alpha}\Omega_{i}^{\alpha}\Omega_{j}^{\alpha}+\sum_{i\in V}\sum_{j\in V}\sum_{\alpha\in S}\sum_{\beta<\alpha}F_{ij}^{\alpha\beta}\Omega_{i}^{\alpha}\Omega_{j}^{\beta}\label{eq:sdss}\,.
\end{equation}
Remarkably, as shown in \cite{Bardella} this allows to introduce the \textbf{Lagrangian} of the system, that is obtained by
simply removing the sum over $\alpha$ in the previous formula.
We interpret the space--like interactions as the \textbf{potential term},
and everything else that couples to the past with the \textbf{kinetic
term}.
A detailed Lagrangian formulation is given in Section 4.1.3 of Bardella
et al. 2024 \cite{Bardella}.
Notice that we interpret as space--like only those parameters with
$\beta=\alpha$, i.e., acting within the same time slice.
The final approximations to obtain the simplified action of \cite{Bardella} are what we call "local memory" and "bi--stationarity", respectively.
By local memory we mean that the columns of the kernel $\Omega$ form an adapted progression of Markov blankets \cite{Franchini2024,Friston3} where there are no interactions between different neurons at different times, i.e., the interaction at different times can be only a self--interaction (like a memory of the past).
In other words, we postulate that the interactions between different neurons are fast enough to be considered purely space--like, which looks like classical (non--relativistic) time evolution.
This is why in \cite{Bardella} this step is called "non--relativistic" truncation: 
\begin{equation}
F_{ij}^{\alpha\beta}=0,\ \ \ \beta<\alpha,\ \ \ j\neq i\,.
\end{equation}
Under this assumption the action can be simplified into \cite{Bardella,iScience2024}
\begin{equation}
\mathscr{A}(\Omega|\,F,I)=\sum_{i\in V}\sum_{\alpha\in S}I_{i}^{\alpha}\Omega_{i}^{\alpha}
+\sum_{i\in V}\sum_{j\in V}\sum_{\alpha\in S}F_{ij}^{\alpha\alpha}\,\Omega_{i}^{\alpha}\Omega_{j}^{\alpha}+\sum_{\alpha\in S}\sum_{\beta\in S}\sum_{i\in V}F_{i\,i}^{\alpha\beta}\Omega_{i}^{\alpha}\Omega_{i}^{\beta}\label{eq:ffd}\,.
\end{equation}
Finally, if the terms with $i\neq j$ are stationary in
$\alpha$ and those with $\alpha\neq\beta$ are stationary in $i$ \cite{Bardella,iScience2024}:
\begin{equation}
F_{ij}^{\alpha\alpha}=A_{ij},\ \ \ F_{ii}^{\alpha\beta}=B^{\alpha\beta},
\end{equation}
which means that the neurons are assumed to be all of the same kind
and that the synaptic couplings don't change in the considered time interval, we call this bi--stationarity of the couplings.
The action is therefore reduced to the following simplified expression:
\begin{equation}
\mathscr{A}(\Omega|\,A,B,I)=\sum_{i\in V}\sum_{\alpha\in S}I_{i}^{\alpha}\Omega_{i}^{\alpha}
+\sum_{i\in V}\sum_{j\in V}A_{ij}\sum_{\alpha\in S}\Omega_{i}^{\alpha}\Omega_{j}^{\alpha}+\sum_{\alpha\in S}\sum_{\beta\in S}B^{\alpha\beta}\sum_{i\in V}\Omega_{i}^{\alpha}\Omega_{i}^{\beta}\label{eq:dfgfdgd}
\end{equation}
Let now introduce the \textbf{space correlation matrix} and the \textbf{time
correlation matrix}
\begin{equation}
\Phi:=\frac{\Omega\,\Omega^{\dagger}}{T},\ \ \ \Pi:=\frac{\Omega^{\dagger}\Omega\,}{N}\,.
\end{equation}
The information is thus coded in three observables $\langle\Phi\rangle_{\mu}$, $\langle\Pi\rangle_{\mu}$ and $\langle\Omega\rangle_{\mu}$,
\begin{equation}
\{A,B,I\}\rightleftharpoons\{\langle\Phi\rangle_{\mu},\langle\Pi\rangle_{\mu},\langle\Omega\rangle_{\mu}\}\,.
\end{equation}
We called this triple \textquotedbl\textbf{hypermatrix}\textquotedbl{},
because the three observables could be arranged into a single matrix like in Figure 2.3 of \cite{Franchini2023}.
The action can be rewritten in its final form \cite{Bardella,iScience2024}:
\begin{equation}
\mathscr{A}(\Omega|\,A,B,I)=\sum_{i\in V}\sum_{\alpha\in S}I_{i}^{\alpha}\Omega_{i}^{\alpha}+T\sum_{i\in V}\sum_{j\in V}A_{ij}\Phi_{ij}+N\sum_{\alpha\in S}\sum_{\beta\in S}B^{\alpha\beta}\Pi^{\alpha\beta}\,.
\end{equation}
It can be shown that this simplified action includes the model used
by Schneidman et al.
2006 \cite{Schneidman}, the PCA and the Hopfield model as special cases \cite{Bardella}.
Finally, we introduce the covariance matrices:
\begin{equation}
\langle\delta\Phi\rangle_{\mu}:=\langle\Phi\rangle_{\mu}-\frac{\langle\Omega\rangle_{\mu}\langle\Omega\rangle_{\mu}^{\dagger}}{T}\, ,\ \ \ 
\langle\delta\Pi\rangle_{\mu}:=\langle\Pi\rangle_{\mu}-\frac{\langle\Omega\rangle_{\mu}^{\dagger}\langle\Omega\rangle_{\mu}}{N}.
\end{equation}
The parameters $A$, $B$ can be inferred from these matrices alone
\begin{equation}
\left\{ A,B\right\} \rightleftharpoons\{\langle\delta\Phi\rangle_{\mu},\langle\delta\Pi\rangle_{\mu}\}.
\end{equation}
Remarkably after these approximations the number of parameters is reduced
from $N^{2}T^{2}$ to $N^{2}+T^{2}$ (actually half due to symmetry) that is a significant gain.
\section{Example: neural recordings with Utah 96}

\noindent{}
The Utah 96 BCI \cite{Utah}
is a silicon--based microelectrode array in the form of a rectangular
or square grid in a $10\,x10$ pattern.
The electrodes
are electrically insulated from neighboring electrodes by a glass
moat surrounding the base, while the tips are coated with platinum,
to facilitate charge transfer into the nerve tissue.
The electrode stems are insulated with silicon nitride.

In \cite{Bardella,iScience2024} we used the \textbf{columnar model} \cite{Jones,Casanova}
as reference for the anatomical organization of the neocortex.
The
neurons are grouped in a two--dimensional lattice of cortical columns, a system in 2+$\epsilon$ dimensions that in turn constitute the
cortex structures and areas (see Figures 3 and 5 of \cite{Bardella}).
The interface is designed to take individual columns with
each needle, at a distance enough to avoid self--interaction terms,
so we can assume that, apart from systematic errors, sensor degradation
etc. the data can be identified with a decimated version of the kernel
of the columnar activities, defined in Section 5.1 of \cite{Bardella} (see \cite{Bardella,Pani} for the details of the experiment).
To model the the interelectrode pitch we therefore apply a renormalization by decimation.
The measurement points can be organized
at the columnar height $z$ (around the inner Baillager band for premotor cortices recordings in monkeys, Figures 3 and 5 of \cite{Bardella}), in a sub--lattice $x'y'\in\mathbb{L}'_{2}$
whose step is much larger than the diameter of the individual
cortical column, so that the activity recorded at different channels belongs
to well spaced columns \cite{Utah}.
In the end we get
the decimated kernel: 
\begin{equation}
\hat{\Omega}:=\{\Omega_{x'y'z}^{\alpha}\in\Gamma:\,x'y'\in\mathbb{L}'_{2},\,\alpha\in S\}\label{eq:experikernel}
\end{equation}
that in \cite{Bardella} we called \textbf{channels kernel}, since it describes the on/off
activity recorded by the channels of the probe.
Examples from actual experiments are in Figures 1, 2 and 7--17 of \cite{Bardella}.
In case of biological neurons observed for a few hundreds of milliseconds (where we expect to have a quenched synapse chemistry) it is expected that the system is adequately captured by the the simplified action and that the parameters $A_{ij}$ and $B^{\alpha\beta}$ can be deduced from the average covariance matrices.
Let us call $\Lambda_{ij}$ the adjacency matrix of the network, and assume that the number of nearest neighbors of each neuron is of the order $N^{\nu}$, where $\nu=1$ is for densely connected models, $\nu<1$ is sparse, and $\nu=0$ is with a finite connectivity.
Following \cite{Schneidman} and the experimental findings in \cite{Bardella} we propose the following approximate form for the matrix element of $A$:
\begin{equation}
A_{ij}=D_{ij}\Lambda_{ij}  +J_{0}\,\Lambda_{ij} N^{-\nu}+J_{ij}\,\Lambda_{ij} N^{-\nu/2}
\end{equation}
where $J_{0}$ is constant, the $J_{ij}$ are Normal i.i.d.
instances and $D_{ij}$ has finite connectivity. We remark that although \cite{Schneidman} assumes a fully conncted $\Lambda_{ij}$ in \cite{Braitenberg} von Braitenberg proposed that the scaling of the fluctuations should have an exponent $\nu=1/2$ (and not $\nu=1$ like in the Sherrington--Kirkpatrick model) suggesting that $\Lambda_{ij}$ is a sparse matrix.
Also, notice that the normalizations where chosen in such a way that the thermodynamic limit (for $N\rightarrow\infty$) exists but this is non--trivial for a biological network and should be tested experimentally (see e.g. Figure 4 of \cite{Bardella}).
Concerning the $B$ matrix, for single neuron recordings like in \cite{Bardella, Pani} its shape could be deduced from the expected dynamics of the single neurons.
This is controlled by the full refractory period $\tau$ so that the entries of $B$ are expected to peak around the line $\alpha - \beta=\tau$.
As experimentally confirmed in \cite{Bardella} from the shape of the time covariance, we can conclude that due to the presence of a relative refractory period \cite{Bardella} coupled to the incoming activity (and therefore to the input $I$) the $\tau$ parameter should be taken time dependent.
We therefore propose the following approximate shape for $B$:
\begin{equation}
    B^{\alpha\beta}=f(\,\alpha-\beta\,,\,\tau^\alpha),
\ \ \ 
    1/\tau^\alpha := a+ \frac{b}{N}\sum_{i\in V}I ^\alpha _i 
\end{equation}
where $a$ and $b$ are positive parameters, and the $f(x,y)$ is a function peaked (in $x$) around the value $y$.
This can capture the non--stationary behavior observed in the time covariance matrices of Figure 14 of \cite{Bardella}, where the forbidden band around the diagonal of the experimental time covariance $\langle\delta\hat{\Pi}\rangle_\mu$ visibly thins out for those times at which the average activity increases.
\section{Conclusions and perspectives}

\noindent{}
In \cite{Bardella,iScience2024} we showed that applying lattice methods from elementary particle theory
to real neurons is possible and fruitful, although the transition
to systematically thinking in this theoretical framework will still require
substantial work.
However, given the advanced state of LFTs and their
large range of applicability, \textbf{knowledge exchange} with the neuroscience
would be beneficial for the theoretical development of the latter
in the near future, and for both in the long run.
While studying data like \cite{Bardella} is a good starting point to develop further capabilities, the network recorded in \cite{Bardella} cannot engage the time features of our LFT framework in a substantial way.
Although the observed neuron dynamics is clearly not stationary due to the relative component of the refractory period (see the non--stationary time covariance in the upper panel of Figure 14 of \cite{Bardella}), it is however quite simply shaped, and we expect that it will only slightly influence the space covariance and the reconstructed space couplings.
We expect that more complex recordings should be analyzed to fully exploit the LFT framework also in the time domain.
For example, non--trivial time covariance matrices could be obtained from long recordings, where some genuine memory effect may well be observed.
We remark that some of these data are quite heavy, and even their manipulation and visualization could become challenging in some cases.
The fact that heavy data are involved is certainly another matter that would require the expertise of the LFT community.
Examples of such datasets can be found in \cite{Manley,Qi,IBL,Bardella2016,DeathKernel}. 

Another interesting direction would be to apply the Neural LFT to some trained neural network by identifying the layers of the network with the columns of the kernel \cite{Bardella,iScience2024,Franchini2023}.
In particular, it would be interesting to look at the hypermatrix of the kernel of activations, possibly averaged on different inputs.
This is still largely unexplored, although an interesting precursor is given in \cite{BardellaTransformers}.
Finally, there are simple models of binary LFT dynamics with long memory, like the Elephant Random Walk \cite{FranchiniHLS3} or the HLS model \cite{FranchiniHLS1,FranchiniHLS2,FranchiniHLS}, that can be analytically solved and eventually used to benchmark the reconstruction of the couplings and other analytical tests.
\section{Acknowledgments}
\noindent{}
This research was partially supported by Sapienza University of Rome, grant PH11715C823A9528 and RM12117A8AD27DB1 Sapienza RM123188F7B71E57.
We acknowledge a contribution from
the Italian National Recovery and Resilience Plan (NRRP) M4C2, funded
by the European Union Next Generation EU (Project IR0000011, CUP
B51E22000150006, EBRAINS Italy).
We thank Stefano Ferraina (Sapienza) for bringing to our attention the reference \cite{Braitenberg}.
\noindent{}

\end{document}